\documentstyle[12pt]{article}
\input{epsfig.sty}
\newcommand{\beq}{\begin{eqnarray}}
\newcommand{\eeq}{\end{eqnarray}}
\begin{document}
\title{Upsilon Production In pp Collisions For Forward Rapidities At LHC}
\author{Leonard S. Kisslinger\\
Department of Physics, Carnegie Mellon University, Pittsburgh PA 15213 USA.\\
Debasish Das\\
Saha Institute of Nuclear Physics,1/AF, Bidhan Nagar, Kolkata 700064, INDIA.}
\date{}
\maketitle
\noindent
PACS Indices:12.38.Aw,13.60.Le,14.40.Lb,14.40Nd
\vspace{1mm}
\begin{abstract}
  This is a continuation of recent studies of $\Upsilon(nS)$ production at 
the LHC in pp collisions. Our previous studies were for rapidity y=-1 to 1
for the CMS detector, while the present study is for y=2.5 to 4.0 at the LHC.
\newline
\end{abstract}
\vspace{1mm} 
{\it Keywords~:~}Quark-Gluon Plasma; upsilon production; forward rapidity; 
standard model; mixed hybrid theory. 

\section{Introduction}

  Recently we have calculated $\Upsilon(nS)$ production in p-p collisions for 
parameters at Fermilab\cite{kmm11} and the LHC\cite{k12}. We used 
the color octet model, which was shown to dominate color singlet production
of heavy quark states in p-p collisions\cite{nlc03,cln04}, with the
treatment formulated by Nayak and Smith\cite{ns06}. 
One main objective of this work is to study how our mixed hybrid 
theory\cite{lsk09} differs from the standard quark model for $\Upsilon(nS)$ 
production. In this theory the $\Upsilon(1S)$ and $\Upsilon(2S)$ states are
conventional $b\bar{b}$ states, while the $\Upsilon(3S)$ approximately 50\% 
standard $b\bar{b}$ and 50\% hybrid.

  Another motivation for these studies is related to possible detection of
Quark-Gluon Plasma (QGP), which is believed to be the nature of the universe 
when the temperature was higher than about 150 MeV, via relativistic heavy
ion collisions. This is being studied at the Relativistic Heavy Ion 
Collider(RHIC), BNL, and the LHC. Since hybrid mesons have active glue, the 
production ratio of $\Upsilon(3S)$ to $\Upsilon(1S)$ production could provide 
evidence for the creation of the QGP. Our present research on $\Upsilon(3S)$ 
vs $\Upsilon(1S)$ via pp collisions at the LHC is an important preliminary to 
this future research. 

\subsection{Mixed heavy quark hybrid and pp collisions}

  One possible signal of QGP is the production of heavy quark states in 
relativistic heavy-ion collisions.
Studies of heavy quark state production in proton-proton (pp) collisions
as a preliminary to relativistic heavy-ion collisions have been carried 
out~\cite{bc96,nlc03,cln04,ns06}.
An important observation concerning the nature of heavy quark charmonium and
bottomonium states are anomalies: a much larger production of $\Psi'(2S)$ in 
high energy collisions than standard model predictions~\cite{cdf}, and the
anomalous production of sigmas in the decay of $\Upsilon(3S)$ to  
$\Upsilon(1S)$~\cite{vogel}. A solution of these anomalies was found in
the mixed hybrid theory~\cite{lsk09}. The $\Psi'(2S)$ state was found to be
\beq
\label{1}
        |\Psi'(2S)>&=& \alpha |c\bar{c}(2S)>+\sqrt{1-\alpha^2}|c\bar{c}g(2S)> 
\; ,
\eeq
where $c$ is a charm quark, and the $\Upsilon(3S)$ state was found to have 
the form
\beq
\label{2}
    |\Upsilon(3S)>&=& \alpha |b\bar{b}(3S)>+\sqrt{1-\alpha^2}|b\bar{b}g(3S)> 
\; ,
\eeq
where $b$ is a bottom quark and $\alpha = -0.7 \pm 0.1$. This means that 
these states have approximately a 50\% probability of being a standard 
quark-antiquark, $|q\bar{q}>$, meson, and a 50\% probability of a hybrid, 
$|q\bar{q}g>$ with the $|q\bar{q}>$ a color octet and $g$
an active gluon. With a valence gluon it would be natural for these hybrid
states to be produced during the creation of a dense QGP.

  Using this mixed hybrid theory, it was shown in our recent work~\cite{kmm11},
upon which the present work is based, that the ratios of $\Psi'(2S)/(J/\Psi)$
and $\Upsilon(3S)/\Upsilon(1S)$ agreed with experimental results, while the
standard model for the $\Psi'(2S)$ and $\Upsilon(3S)$ did not.

\subsection{Heavy quarkonium state production at forward rapidities}

At the LHC the parton distribution functions of the 
nucleon can be studied in pp collisions, and in pA and AA collisions their 
modifications in 
the nucleus for very low values of momentum fraction (Bjorken $x$).
The capabilities to measure charm and beauty particles in the forward 
rapidity region ($|y|\simeq4$) gives access to the regime of $x\sim10^{-6}$ 
in LHC experiments, such as A Large Ion 
Collider Experiment(ALICE)~\cite{dd10,dd11,dd11z,dd12,dd13} and Large Hadron Collider 
beauty(LHCb)~\cite{LHCb12}

Using the ALICE Muon Spectrometer at the LHC, charm and beauty particles can be 
measured in the forward rapidity region (2.5 $<$ y $<$ 4)  via 
its di-muon decay. The main goal of the ALICE Muon Spectrometer physics 
program~\cite{dd10,dd11,dd11z} is based on the measurement
of heavy-flavor production in forward rapidity region for pp, pA and AA
collisions at LHC energies. For the relativistic heavy-ion collisions the 
dependence with the collision centrality and 
the reaction plane can be studied. For AA collisions $J/\Psi$ suppression
is a signature of the QGP~\cite{ms86}, and $\Upsilon$ 
states are  expected to dissociate at a
higher temperature~\cite{ddups1,ddups2} than the other quarkonium states,
thus proving to be a more effective thermometer of the
system. With the $\Psi'(2S)$ and $\Upsilon(3S)$ states having active gluons, 
measurement of their production can be a test of the creation of QGP.

\section{Differential rapidity distribution for $\Upsilon(nS)$ 
production}

  In our recent work on $\Upsilon(nS)$ production via proton-proton (pp)
collisions we followed the formalism of Refs[~\cite{nlc03,cln04,ns06} for
E=energy=$\sqrt{s}$ = 2.76 TeV~\cite{kmm11} and 7.0 TeV~\cite{k12}, with 
the rapidity variable y = -1 to +1, for the Compact Muon Solenoid(CMS) 
detector. 
In the present work we calculate $\Upsilon(nS)$ production for the 
acceptance(y=2.5 to 4.0) of the Muon Spectrometer at the LHC. 
The differential rapidity distribution 
for $\lambda=0$ (dominant for $\Upsilon(nS)$ production) is given 
by~\cite{kmm11} ($A_\Upsilon=1.12\times 10^{-7},1.73\times 10^{-8}$ nb, 
for $\sqrt{s}$ = 2.76, 7.0 TeV)
\beq
\label{3}
      \frac{d \sigma_{pp\rightarrow \Phi(\lambda=0)}}{dy} &=& 
     A_\Upsilon \frac{1}{x(y)} f_g(x(y),2m)f_g(a/x(y),2m) \frac{dx}{dy} \; ,
\eeq 

\beq
\label{4}
   x(y) &=&  \left[\frac{m}{2E}(\exp{y}-\exp{(-y)})+\sqrt{(\frac{m}{E}
(\exp{y}-\exp{(-y)}))^2 +4a}\right] \nonumber \\
  \frac{d x(y)}{d y} &=&\frac{m}{2E}(\exp{y}+\exp{(-y)})\left[1. + 
\frac{\frac{m}{E}(\exp{y}-\exp{(-y)})}{\sqrt{(\frac{m}{E} 
(\exp{y}-\exp{(-y)}))^2 +4a}}\right] \; ,
\eeq
with $a= 4m^2/s$, $m=5.$ GeV, and $f_g$ the gluonic distribution function.

  Using Eqs(\ref{3},\ref{4}) and parameters given in Ref~\cite{kmm11} we obtain
the results for $\Upsilon(1S)$ and $\Upsilon(3S)$ production shown in Fig. 1
and Fig. 2 at 2.76 TeV and 7.0 TeV in pp collisions for $2.5 \leq y \leq 4.0$.
\vspace{1cm}

\begin{figure}[ht]
\begin{center}
\epsfig{file=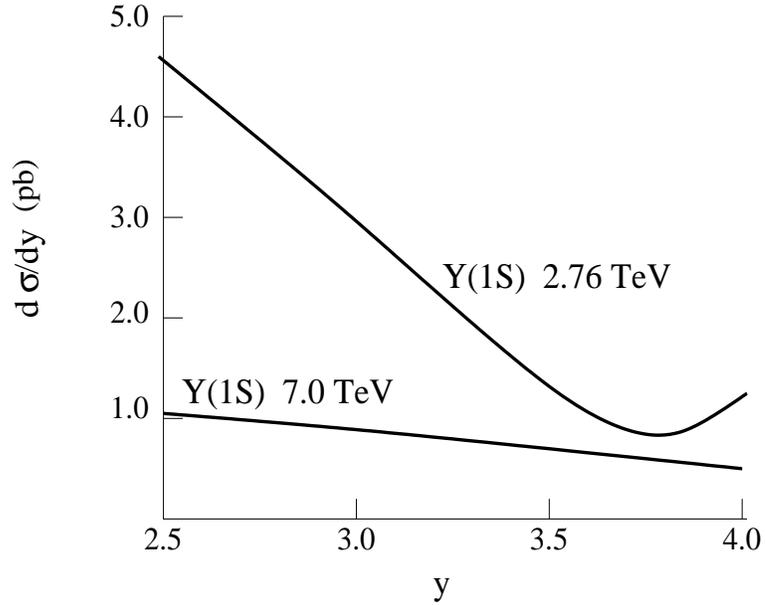,height=8cm,width=10cm}
\caption{d$\sigma$/dy for pp collisions at $\sqrt{s}$ = 2.76 and 7.0 producing 
$\Upsilon(1S)$.}
\label{Figure 1}
\end{center}
\end{figure} 
\vspace{2cm}

  Although the units in Figs. 1, 2 are in pb, the actual magnitude is 
uncertain due to the normalization of the state. The overall magnitude 
and rapidity dependence of the differential rapidity distribution, however, 
provides satisfactory estimates at forward rapidities for LHC experiments.

\newpage

\vspace{1cm}

\begin{figure}[ht]
\begin{center}
\epsfig{file=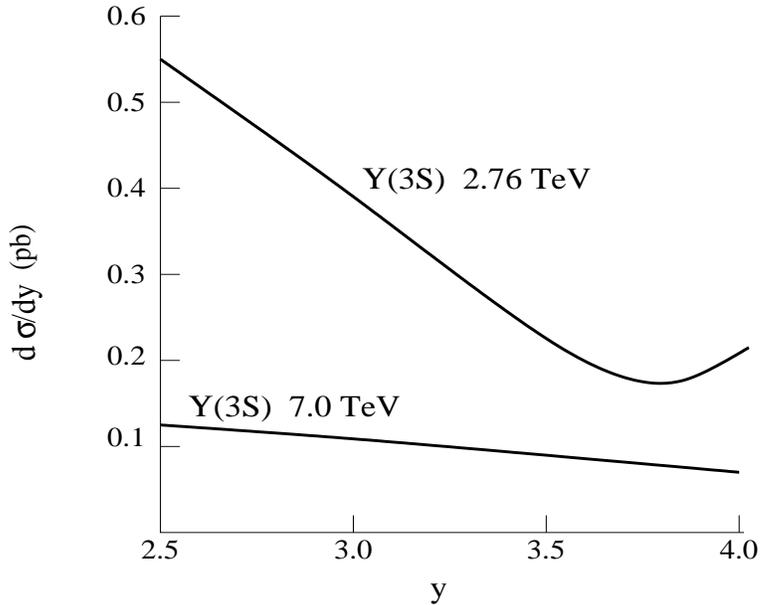,height=8cm,width=10cm}
\caption{d$\sigma$/dy for pp collisions at $\sqrt{s}$ = 2.76 and 7.0 TeV 
producing $\Upsilon(3S)$.}
\label{Figure 2}
\end{center}
\end{figure}

\subsection{Ratios of $\Upsilon(2S)$ and $\Upsilon(3S)$ to 
$\Upsilon(1S)$} 

The ratios of $\Upsilon(2S)$ and $\Upsilon(3S)$ for the standard model and
for the mixed hybrid model~\cite{lsk09}, upon which the present work is based,
are given by the premise that the $\Upsilon(3S)$ is 50\% standard and 50\% 
hybrid, as shown in Eq.(\ref{2}). One finds (see Ref~\cite{kmm11} for a 
detailed discussion) that 

\beq
\label{5}
      [\sigma(\Upsilon(2S))/\sigma(\Upsilon(1S))]_{mixed} &=&
       [\sigma(\Upsilon(2S))/\sigma(\Upsilon(1S))]_{standard} \simeq 0.27
 \nonumber 
\eeq
\beq
  [\sigma(\Upsilon(3S))/\sigma(\Upsilon(1S))]_{mixed} &\simeq&
    2.5 \times [\sigma(\Upsilon(3S))/\sigma(\Upsilon(1S))]_{standard} 
\simeq 0.1 \; .
\eeq
These are the same as those given in Ref~\cite{k12}, and are consistent with 
CMS measurements. 

Note that recent measurements by the LHCb experiment 
of the $\Upsilon$ production in pp collisions at $\sqrt{s}$ = 7 
TeV~\cite{LHCb12}
in the rapidity range $2.0 < y < 4.5$ found the cross sections times the 
branching fractions to $\mu^+ \mu^-$ are $\Upsilon(2S)/\Upsilon(1S) \simeq 
0.25$, in agreement with the standard and mixed hybrid theories, while 
$\Upsilon(3S)/\Upsilon(1S) \simeq 0.12$, in agreement with the mixed hybrid 
theory, but in disagreement with the standard model.

\section{Conclusions}

  We expect that our results for the rapidity dependence of
d$\sigma$/dy shown in Figs. 1 and 2 can be  useful in the forward rapidity 
bottomonia measurements at the LHC. Also, the ratios of the 
production of $\Upsilon(3S)$ to
$\Upsilon(1S)$ will be both a test of the validity of the mixed hybrid
theory, and should also be a guide for LHC experiments. These experimental and 
theoretical studies will serve as a basis for tests of the creation of
the QGP in ultra-relativistic heavy-ion collision experiments at the LHC.

\Large{{\bf Acknowledgements}}

\vspace{5mm}
\normalsize
This work was supported in part by a grant from the Pittsburgh Foundation. 
Author DD acknowledges the facilities 
of Saha Institute of Nuclear Physics, Kolkata, India.

\end{document}